\documentclass[amsmath,amssymb,aps,longbibliography,pre,showpacs,superscriptaddress,twocolumn]{revtex4-1}

\usepackage{graphicx} 
\usepackage{hyperref} 
\usepackage{dcolumn} 

\begin{document}

\preprint{EH11219}

\title{Velocity and energy distributions in microcanonical ensembles of hard
spheres}

\author{Enrico Scalas}
\email{e.scalas@sussex.ac.uk}
\homepage{www.sussex.ac.uk/profiles/330303}
\affiliation{School of Mathematical and Physical Sciences,
University of Sussex, Falmer, Brighton BN1 9RH, UK}
\affiliation{Basque Center for Applied Mathematics, Alameda de Mazarredo 14,
48009 Bilbao, Basque Country, Spain}
\author{Adrian T.~Gabriel}
\email{adriangabriel@gmx.de}
\affiliation{Department of Chemistry and WZMW, 
Philipps-University Marburg, 35032 Marburg, Germany}
\author{Edgar Martin}
\email{emartin@gmx.de}
\affiliation{Department of Chemistry and WZMW, 
Philipps-University Marburg, 35032 Marburg, Germany}
\author{Guido Germano}
\email{g.germano@ucl.ac.uk}
\homepage{www.cs.ucl.ac.uk/staff/g.germano}
\affiliation{Department of Computer Science, 
University College London, Gower Street, London WC1E 6BT, UK}
\affiliation{Systemic Risk Centre, London School of Economics and Political
Science, Houghton Street, London WC2A 2AE, UK}

\date{\today}

\begin{abstract}
In a microcanonical ensemble (constant $NVE$, hard reflecting walls)
and in a molecular dynamics ensemble (constant $NVE\mathbf{PG}$, periodic
boundary conditions) with a number $N$ of smooth elastic hard spheres in
a $d$-dimensional volume $V$ having a total energy $E$, a total momentum
$\mathbf{P}$, and an overall center of mass position $\mathbf{G}$,
the individual velocity components, velocity moduli, and energies have
transformed beta distributions with different arguments and shape parameters
depending on $d$, $N$, $E$, the boundary conditions, and possible symmetries in
the initial conditions. This can be shown marginalizing the joint distribution
of individual energies, which is a symmetric Dirichlet distribution. In the
thermodynamic limit the beta distributions converge to gamma distributions with
different arguments and shape or scale parameters, corresponding respectively
to the Gaussian, i.e., Maxwell-Boltzmann, Maxwell, and Boltzmann or
Boltzmann-Gibbs distribution. These analytical results agree with molecular
dynamics and Monte Carlo simulations with different numbers of hard disks
or spheres and hard reflecting walls or periodic boundary conditions.
The agreement is perfect with our Monte Carlo algorithm, which acts only
on velocities independently of positions with the collision versor sampled
uniformly on a unit half sphere in $d$ dimensions, while slight deviations
appear with our molecular dynamics simulations for the smallest values of $N$.
\end{abstract}


\maketitle

\section{Introduction}

The problem of the velocity distribution in a gas of hard spheres was discussed
in a paper published by Maxwell in 1860 \cite{Maxwell1860a}. Maxwell obtained
the velocity distribution by assuming independence of the three components
of velocity and rotational invariance of the joint distribution. The only
distribution satisfying the functional equation
\begin{align}
\label{eq:rotinv}
f_\mathbf{v}(x_1,x_2,x_3) &= \Phi(x_1^2 + x_2^2 + x_3^2) \nonumber \\
&= f_{v_1}(x_1) f_{v_2}(x_2) f_{v_3}(x_3) 
\end{align}
has factors of the form
\begin{equation}
\label{eq:clt1}
f_{v_\alpha}(x) = A \exp(-B x^2),
\end{equation}
$\alpha = 1, 2, 3$.
This simple heuristic derivation can still be found in modern textbooks in
statistical physics or physical chemistry \cite{Atkins1978}, but
generalizations of Maxwell's method appeared earlier in the physical literature
\cite{Maizlish1922}.

In 1867, Maxwell \cite{Maxwell1867} became aware that Eq.~(\ref{eq:clt1})
should appear as a stationary solution for the dynamics of the gas and
introduced the assumption of molecular chaos, according to which the velocities
of two colliding molecules are uncorrelated and independent of their positions.
This concept was later called \textit{Sto{\ss}zahlansatz} or collision number
hypothesis by Boltzmann. It led to a more detailed study of molecular
collisions and to kinetic equations whose stationary solutions coincide with
Maxwell's original distribution; see Refs.~\cite{Desvillettes1999,Graham1999,
Fournier2001a,Fournier2001b,Fournier2001c} for a modern mathematical approach
to kinetic equations. This route was followed by Boltzmann, who obtained the
velocity distribution in a more general way in a series of papers published
between 1868 and 1871 \cite{Boltzmann1868,Boltzmann1871a,Boltzmann1871b}.
Based on the \textit{Sto{\ss}zahlansatz}, Boltzmann could prove that Maxwell's
distribution is stationary. These results are summarized in Tolman's book
\cite{Tolman1938} and in the first chapter of ter Haar's book
\cite{terHaar1954}. The \textit{Sto{\ss}zahlansatz} provides an answer
to Loschmidt's \textit{Umkehreinwand} or reversibility paradox of 1876
\cite{Loschmidt1876}, which questions how time-reversible microscopic dynamics
can lead to time-irreversible results like the increase of the entropy of a
gas as stated by Boltzmann's H-theorem of 1872 \cite{Boltzmann1872}: it was
understood only later that the microscopic time-reversibility of Newton's
equations of motion was effectively destroyed by Boltzmann's use of the
\textit{Sto{\ss}zahlansatz} in his calculations. Boozer \cite{Boozer2011}
provided a recent analysis of the role of the assumption of molecular chaos
to obtain time-asymmetric results like Boltzmann's H-theorem and the Boltzmann
transport equations by performing computer simulations of a one-dimensional
system of two interlieved molecular species with mass ratios 1:2.

Boltzmann and Maxwell were not working in isolation and were aware of their
respective works. In his 1872 paper, Boltzmann often quotes Maxwell
\cite{Boltzmann1872}. In 1873, Maxwell wrote to his correspondent Tait
\cite{Maxwell1873}:
\begin{quote}
By the study of Boltzmann I have been unable to understand him. He could not
understand me on account of my shortness, and his length was and is an equal
stumbling block to me.
\end{quote}
More details on the relationship between Maxwell and Boltzmann and on the
influence of Maxwell on Boltzmann's thought have been collected by
Uffink~\cite{Uffink2008}.

Tolman's analysis of classical binary collisions for hard spheres led to rate
equations which can be interpreted as transition probabilities for a Markov
chain after proper normalization. The interested reader can consult chapter V
of Tolman's classic book \cite{Tolman1938}, in particular the discussion around
Eq.~(45.3) on page 129. The connection with Markov chains was made explicit by
Costantini and Garibaldi \cite{Costantini1997,Costantini1998}, who used a model
due to Brillouin \cite{Brillouin1927}. Before Costantini and Garibaldi, Penrose
suggested that a Markovian hypothesis could justify the use of standard
statistical mechanical tools \cite{Penrose1970}. According to our
interpretation of Penrose, due to the limits in human knowledge naturally
leading to coarse graining, systems of many interacting particles effectively
behave as Markov chains. Moreover, the possible number of states of such a
chain is finite even if very large, therefore only the theory of finite Markov
chains is useful. Statistical equilibrium is reached when the system states
obey the equilibrium distribution of the finite Markov chain; this equilibrium
distribution exists, is unique and coincides with the stationary distribution
if the chain is irreducible and aperiodic. This point of view is also known as
Markovianism. Indeed, in a recent paper on the Ehrenfest urn, or dogs and fleas
model, we showed that, after appropriate coarse graining, a Markov chain well
approximates the behavior of a realistic model for a fluid \cite{Scalas2007}.

Here, we study the velocity distribution in a system of $N$ smooth elastic
hard spheres in $d$ dimensions. Even if the evolution of the system is
deterministic, we can consider the velocity components of each particle as
random variables. We do not consider a finitary \cite{Garibaldi2010} version
of the model by discretizing velocities, but keep them as real variables.
Then a heuristic justification of Eq.~(\ref{eq:clt1}) can be based on the
central limit theorem (CLT). Here is the argument. Following Maxwell's idea,
one can consider the velocity components of each particle independent from
each other. Further assuming that velocity jumps after collisions are
independent and identically distributed random variables, one obtains for
the velocity component $\alpha$ of a particle $i$ at time $t$
\begin{equation}
\label{eq:randomwalk}
v_{i\alpha}(t) = v_{i\alpha}(0) + \sum_{j=1}^{n(t)} \Delta v_{i\alpha,j},
\end{equation}
where $n(t)$ is the number of collisions for that particle up to time $t$
and $\Delta v_{i\alpha,j}$ is the change in velocity at collision $j$.
If the hypotheses stated above are valid, Eq.~(\ref{eq:randomwalk}) defines
a continuous-time random walk and the distribution $f_{v_{i\alpha}}(x,t)$
approaches a normal distribution for large $t$ as a consequence of the CLT.
Unfortunately this argument is only approximately true in the case of large
systems and false for smaller systems.

In Sec.~\ref{sec:theory} we obtain the theoretical probability density
functions of the individual energies, velocity moduli and velocity components,
starting from the fundamental uniform distribution law in phase space. In
Sec.~\ref{sec:md} we present the molecular dynamics method used to simulate
hard spheres. Interestingly, the same distributions can be reproduced by a
simple Monte Carlo stochastic model introduced in Sec.~\ref{sec:mc}. The
numerical results are presented in Sec.~\ref{sec:results} together with some
statistical goodness-of-fit tests. Indeed, it turns out that an equilibrium
distribution of the velocity components seems to be reached already for $N = 2$
particles and without using any coarse graining. When $N$ grows the equilibrium
distribution approaches the normal distribution, Eq.~(\ref{eq:clt1}).
A discussion and a summary follow in Sec.~\ref{sec:conclusions}.

\section{Theory}
\label{sec:theory}

We consider a fluid of $N$ hard spheres in $d$ dimensions with the same
diameter $\sigma$ and mass $m$ in a cuboidal box with sides $L_\alpha,\ \alpha
= 1,\dots,d$. The positions $\mathbf{r}_i$, $i = 1,\dots,N$, are confined to
a $d$-dimensional box with volume $V = \prod_{\alpha=1}^d L_\alpha$, i.e.,
each position component $r_{i\alpha}$ can vary in the interval $[-L_\alpha/2,
L_\alpha/2]$. Elastic collisions transfer kinetic energy between the particles,
while the total energy of the system
\begin{equation}
\label{eq:energy}
E = \frac{1}{2} \sum_{i=1}^N m_i v_i^2 = \frac{1}{2} m v^2
\end{equation}
does not change in time, i.e., it is a constant of the motion. Therefore, the
velocities $\mathbf{v}_i$ are confined to the surface of a hypersphere with
radius $R = \sqrt{2E/m}$ given by the constraint that the total energy is $E$,
i.e., each velocity component $v_{i\alpha}$ can vary in the interval $[-R,R]$
with the restriction on the sum of the squares given by Eq.~(\ref{eq:energy}).
In other words, the rescaled positions $\mathbf{q}$ with $q_{i\alpha} =
r_{i\alpha}/L_\alpha$ are confined to the unit hypercube in $dN$ dimensions,
while the rescaled velocity components $\mathbf{u} = \mathbf{v}/R$ are confined
to the surface of the unit hypersphere in $dN$ dimensions defined by the
constraint $u = \sqrt{\mathbf{u \cdot u}} = 1$.

The state of the system is specified by the phase space vector of all
velocities and positions $\boldsymbol{\Gamma} = (\mathbf{v,r})$, i.e.,
by $2dN$ variables: the velocity components $v_{i\alpha}$ and the position
components $r_{i\alpha}$. However, these variables are not independent because
of constraints. For spheres with random velocities and positions confined
in a container with hard reflecting walls, the total energy $E$ is conserved
and thus the degrees of freedom are $g = 2dN-1$; this is the microcanonical
ensemble (constant $NVE$). Periodic boundary conditions conserve also the total
linear momentum
\begin{equation}
\label{eq:momentum}
\mathbf{P} = \sum_{i=1}^{N} m_i \mathbf{v}_i = m \sum_{i=1}^{N} \mathbf{v}_i
\end{equation}
and the generator of Galilean transformations to other inertial frames of
reference
\begin{equation}
\mathbf{G} = \mathbf{P}t - \sum_{i=1}^{N} m_i \mathbf{r}_i
= m\sum_{i=1}^{N}(\mathbf{v}_i t - \mathbf{r}_i),
\end{equation}
where the coordinates $\mathbf{r}_i$ are not reboxed upon a crossing of the
unit cell boundaries (if $\mathbf{P} = \boldsymbol{0}$, $-\mathbf{G}/\big(
\sum_{i=1}^{N}m_i\big) = -\mathbf{G}/(Nm)$ is the position of the center of
mass), and thus the number of independent variables drops to $g = 2d(N-1)-1 =
2dN-2d-1$; this is the molecular dynamics ensemble (constant $NVE\mathbf{PG}$)
\cite{Erpenbeck1977,Ray1999,Shirts2006}. Symmetries in the positions
and velocities may reduce $g$ further; e.g.\ if all components $i$ of
$\boldsymbol{\Gamma}$ are pairwise symmetric with respect to the origin,
with both kinds of boundary conditions this point symmetry will stay on forever
and $g = dN-1$ or $g = 2d(N/2-1)-1 = dN-2d-1$ respectively. For the sake of
simplicity, in presenting the theory we will treat explicitly only the
microcanonical case without symmetries.

Following Khinchin \cite{Khinchin1949}, one can assume as the starting point
of statistical mechanics that the distribution in the accessible portion of
phase space is uniform, although so far this has not been rigorously proved
in general. In our case, the measure of the accessible region of phase space
is the product of the volume of the hypercube $V^N$ times the surface of the
hypersphere with radius $R$ in $dN$ dimensions,
\begin{equation}
\Omega = V^N \frac{2\pi^{dN/2}}{\Gamma(dN/2)} R^{dN-1}.
\end{equation}
Then Khinchin's \textit{Ansatz} that the probability density function (PDF) for
points $(\mathbf{v,r})$ in the permitted region of phase space is uniform leads
to the joint PDF for velocities and positions
\begin{equation}
\label{eq:uniformprobability1}
f_{\mathbf{v,r}}(\mathbf{x,y}) = \frac{1}{\Omega} \mathbf{1}_{\{\mathbf{x} :
x = R\}}(\mathbf{x}) \mathbf{1}_{\{\mathbf{y} : -L_\alpha/2 \leq y_{i\alpha}
\leq L_\alpha/2\}}(\mathbf{y}),
\end{equation}
where $\mathbf{1}_A(\mathbf{x})$ is the indicator function of the set $A$,
\begin{equation}
\mathbf{1}_A(\mathbf{x}) \stackrel{\mathrm{def}}{=}
\begin{cases}
1 & \mathrm{if}\ \mathbf{x} \in A \\
0 & \mathrm{if}\ \mathbf{x} \notin A
\end{cases}.
\end{equation}
As the energy does not depend on positions, one can integrate over the latter,
yielding a uniform PDF for particle velocities on the surface of a hypersphere,
\begin{equation}
\label{eq:uniformprobability2}
f_\mathbf{v}(\mathbf{x}) = \frac{\Gamma(dN/2)}{2\pi^{dN/2}} R^{1-dN}
\mathbf{1}_{\{\mathbf{x} : x = R\}}(\mathbf{x}).
\end{equation}
The marginalization of this joint PDF leads to the distributions of individual
particle energies as well as of velocity moduli and velocity components.
To this purpose, it is convenient to study the relationship between
Eq.~\eqref{eq:uniformprobability2} and the symmetric Dirichlet distribution
with parameter $a$.

The PDF of the $n$-dimensional Dirichlet distribution with parameter vector
$\mathbf{a}$ is
\begin{equation}
\label{eq:dirichlet}
f_\mathbf{X}^\mathrm{D}(\mathbf{x;a}) \stackrel{\mathrm{def}}{=}
\frac{1}{\mathrm{B}(\mathbf{a})} \prod_{i=1}^n x_i^{a_i-1}
\mathbf{1}_S(\mathbf{x});
\end{equation}
it is zero outside the unit simplex
\begin{equation}
S = \left\{\mathbf{x} \in \mathbb{R}^n: \forall i\ x_i \geq 0 \wedge
\sum_{i=1}^n x_i = 1\right\}.
\end{equation}
The normalization factor is the multinomial beta function, which can be defined
through the gamma function,
\begin{equation}
\label{eq:multinomialbeta}
\mathrm{B}(\mathbf{a}) = \frac{\prod_{i=1}^n\Gamma(a_i)}{\Gamma(\sum_{i=1}^n
a_i)}.
\end{equation}
The multinomial beta function is a generalization of the beta function
or Euler integral of the first kind, $\mathrm{B}(x,y) = \int_0^1 t^{x-1}
(1-t)^{y-1} \mathrm{d}t$, which can be expressed through the gamma function or
Euler integral of the second kind, $\Gamma(z) = \int_0^{\infty} e^{-t} t^{z-1}
\mathrm{d}t$, as $\mathrm{B}(x,y) = \Gamma(x)\Gamma(y)/\Gamma(x+y)$. In the
symmetric Dirichlet distribution all elements of the parameter vector
$\mathbf{a}$ have the same value $a_i=a$,
\begin{equation}
\label{eq:symmetric_dirichlet}
f_\mathbf{X}^\mathrm{D}(\mathbf{x};a) \stackrel{\mathrm{def}}{=}
\frac{\Gamma(na)}{[\Gamma(a)]^n} \prod_{i=1}^n x_i^{a-1}
\mathbf{1}_S(\mathbf{x}).
\end{equation}
Notice that $a = 1$ gives the uniform distribution on $S$.

It is convenient to work with dimensionless variables. The rescaling
$u_{i\alpha} = v_{i\alpha}/R$ introduced above gives the PDF
\begin{equation}
\label{eq:uniformprobability3b}
f_\mathbf{u}(\mathbf{x}) = \frac{\Gamma(dN/2)}{2\pi^{dN/2}}
\mathbf{1}_{\{\mathbf{x} : x = 1\}}(\mathbf{x}).
\end{equation}
A second transformation
\begin{equation}
\label{eq:newvariable}
w_{i\alpha} = u_{i\alpha}^2
\end{equation}
leads to a set of $dN$ random variables each one with support in [0,1] and such
that
\begin{equation}
\label{eq:constant_E}
\sum_{i=1}^N \sum_{\alpha = 1}^d w_{i\alpha} = 1.
\end{equation}
The Jacobian for this transformation is
\begin{equation}
\label{eq:transformation}
\left|\frac{\partial\mathbf{u}}{\partial\mathbf{w}}\right|
= \frac{1}{2^{dN}} \prod_{i=1}^N \prod_{\alpha=1}^d w_{i\alpha}^{-1/2}.
\end{equation}
Multiplying it by a factor $2^{dN}$ because each $\pm u_{i\alpha}$ results
in the same $w_{i\alpha}$ and by another factor 2  because of the constraint
given by Eq.~\eqref{eq:constant_E} (for details see Song and Gupta
\cite{Song1997}), and replacing $\sqrt{\pi} = \Gamma(1/2)$, the joint PDF of
the variables $w_{i\alpha}$ can be expressed through the symmetric Dirichlet
PDF with parameter $a=1/2$,
\begin{align}
\label{eq:dirichlet1}
f_\mathbf{w}(\mathbf{x}) &= f_\mathbf{w}^\mathrm{D}(\mathbf{x};1/2)\nonumber\\
&= \frac{\Gamma(dN/2)} {[\Gamma(1/2)]^{dN}} \prod_{i=1}^N \prod_{\alpha=1}^d
x_{i\alpha}^{-1/2} \mathbf{1}_S(\mathbf{x}).
\end{align}
Now the normalized energy per particle
\begin{equation}
\label{eq:energyperparticle}
\varepsilon_i = \frac{E_i}{E} = \frac{m v_i^2}{2E} = \frac{v_i^2}{R^2}
= \sum_{\alpha=1}^d w_{i\alpha}
\end{equation}
is the sum of $d$ variables distributed according to Eq.~(\ref{eq:dirichlet1}).
As a consequence of the aggregation law for Dirichlet distributions, one finds
that the joint PDF of all $\varepsilon_i$ is
\begin{align}
\label{eq:dirichlet2}
f_{\boldsymbol{\varepsilon}} (\mathbf{x})
&= f_{\boldsymbol{\varepsilon}}^\mathrm{D}(\mathbf{x};d/2) \nonumber \\
&= \frac{\Gamma(dN/2)}{[\Gamma(d/2)]^N} \prod_{i=1}^N x_i^{d/2-1}
\mathbf{1}_S(\mathbf{x}).
\end{align}
It is interesting to notice that this is a uniform distribution for $d=2$;
because of this, Boltzmann's 1868 method \cite{Boltzmann1868} works in $d=2$
dimensions, but fails for $d=3$.

The PDF of the normalized energies of single particles can be obtained by a
further marginalization of the symmetric Dirichlet distribution given by
Eq.~\eqref{eq:dirichlet2}, using again the aggregation law. The result is a
beta distribution, whose PDF is
\begin{equation}
f_X^\beta(x;a,b) \stackrel{\textrm{def}}{=}
\frac{1}{\mathrm{B}(a,b)} x^{a-1}(1-x)^{b-1} \mathbf{1}_{[0,1]}(x),
\end{equation}
i.e.\ a Dirichlet distribution, Eq.~(\ref{eq:dirichlet}), with $n = 2$. In
other words, the Dirichlet distribution is a multivariate generalization of the
beta distribution. Our case has the exponents $a=d/2$ and $b=d(N-1)/2$,
\begin{align}
\label{eq:beta1}
f_{\varepsilon_i}(x) =& f_{\varepsilon_i}^\beta\left(x;\frac{d}{2},
\frac{d(N-1)}{2}\right) \nonumber \\
=& \frac{\Gamma(dN/2)}{\Gamma(d/2)\Gamma(d(N-1)/2)} x^{\frac{d}{2}-1}
\nonumber \\
& \times (1-x)^{\frac{d(N-1)}{2}-1} \mathbf{1}_{[0,1]}(x).
\end{align}
The transformation of variables $E_i = E \varepsilon_i$ immediately leads to
the beta-Stacy PDF of particle energies,
\begin{align}
\label{eq:fEi}
f_{E_i}(x) =& f_{E_i}^\beta\left(\frac{x}{E};\frac{d}{2},
\frac{d(N-1)}{2}\right) \frac{\mathrm{d}}{\mathrm{d}x}\frac{x}{E} \nonumber \\
=& \frac{\Gamma(dN/2)}{\Gamma(d/2)\Gamma(d(N-1)/2)}
\left(\frac{x}{E}\right)^{\frac{d}{2}-1} \nonumber \\
& \times \left(1-\frac{x}{E}\right)^{\frac{d(N-1)}{2}-1}
\frac{1}{E} \mathbf{1}_{[0,E]}(x)
\end{align}
for $N > 1$, and $f_{E_i}(x) = \delta(x-E)$ for $N = 1$. This result has been
obtained with a different method, without invoking the Dirichlet and beta
distributions, by Shirts et al.\ \cite[Eq.~(9)]{Shirts2006}. 

In the thermodynamic limit ($N,V,E \to \infty$ with $N/V = \rho =$ constant
and $E/N = \bar{E} =$ constant), Eq.~(\ref{eq:fEi}) converges to a gamma
distribution, as discussed by Garibaldi and Scalas
\cite[pages 121--122]{Garibaldi2010}. The gamma PDF is
\begin{equation}
f_X^\gamma(x;a,b) \stackrel{\textrm{def}}{=} \frac{x^{a-1}}{b^a\Gamma(a)}
\exp\left(-\frac{x}{b}\right) \mathbf{1}_{[0,\infty)}(x).
\end{equation}
A scale parameter $b$ is usually included in the definition of the gamma
distribution, but it can always be set to 1 absorbing it into the argument,
\begin{equation}
f_X^\gamma(x;a,b) = f_X^\gamma\left(\frac{x}{b};a,1\right)\frac{1}{b} \equiv
f_X^\gamma\left(\frac{x}{b};a\right)\frac{1}{b}.
\end{equation}
Coming back to the thermodynamic limit of Eq.~\eqref{eq:fEi} anticipated
above, this is a gamma distribution with shape parameter $a = d/2$ and scale
parameter $b = 2 \bar{E}/(dm) = R^2/(dN)$,
\begin{multline}
f_{E_i}(x) = f_{E_i}^\gamma\left(x;\frac{d}{2},\frac{2\bar{E}}{dm}\right) \\
= \left(\frac{dm}{2\bar{E}}\right)^{d/2} \frac{x^{d/2-1}}{\Gamma(d/2)}
\exp\left(-\frac{dmx}{2\bar{E}}\right) \mathbf{1}_{[0,\infty)}(x),
\end{multline}
which is the familiar Boltzmann or Boltzmann-Gibbs distribution for $d=2$.

The PDF of the velocity moduli, or speeds, of individual particles can be
obtained from $f_{E_i}(x)$ replacing $E_i = mv_i^2/2$. The result is
a transformed beta-Stacy distribution with the same exponents $a = d/2$ and
$b = d(N-1)/2$ as for the energies, but argument $mx^2/(2E) = (x/R)^2$,
\begin{align}
\label{eq:fvi}
f_{v_i}(x) =& f_{v_i}^\beta\left(\frac{x^2}{R^2};\frac{d}{2},\frac{d(N-1)}{2}
\right) \frac{\mathrm{d}}{\mathrm{d}x} \frac{x^2}{R^2}, \nonumber \\
=& \frac{\Gamma(dN/2)}{\Gamma(d/2)\Gamma(d(N-1)/2)}
\frac{2x^{d-1}}{R^d} \nonumber \\
& \times \left(1-\frac{x^2}{R^2}\right)^{\frac{d(N-1)}{2}-1}
\mathbf{1}_{[0,R]}(x)
\end{align}
for $N > 1$, and $f_{v_i}(x) = \delta(x-R)$ for $N = 1$. Also this result has
been obtained with a different method \cite{Shirts2006,Uline2008,Boozer2010}.

In the thermodynamic limit, Eq.~(\ref{eq:fvi}) converges to a transformed
gamma distribution with argument $x^2/2$, shape parameter $a = d/2$ and
scale parameter $b = R^2/(dN) = 2\bar{E}/(dm)$,
\begin{multline}
f_{v_i}(x) = f_{v_i}^\gamma\left(\frac{x^2}{2};\frac{d}{2},\frac{2\bar{E}}{dm}
\right) \frac{\mathrm{d}}{\mathrm{d}x} \frac{x^2}{2} \\
= \left(\frac{dm}{\bar{E}}\right)^\frac{d}{2} \frac{(x/2)^{d-1}}{\Gamma(d/2)}
\exp\left(-\frac{dmx^2}{4\bar{E}}\right) \mathbf{1}_{[0,\infty)}(x),
\end{multline}
which is the familiar Maxwell distribution for $d = 3$.

The transformation from hyperspherical to cartesian coordinates 
$v_i^2=\sum_{\alpha=1}^d v_{i\alpha}^2$ and 
$(2\pi^{d/2}/\Gamma(d/2))v_i^{d-1} \mathrm{d}v_i = \prod_{\alpha=1}^d
\mathrm{d} v_{i\alpha}$ leads from Eq.~\eqref{eq:fvi} to the PDF
$f_{\mathbf{v}_i}(\mathbf{x})$ of the single-particle velocity vectors
\begin{equation}
f_{\mathbf{v}_i}(\mathbf{x}) = \frac{\Gamma(dN/2)}{\Gamma(d(N-1)/2)}
\left(1-\frac{x^2}{R^2}\right)^{\frac{d(N-1)}{2}-1}
\frac{\mathbf{1}_{\{\mathbf{x} : x = R\}}(\mathbf{x})}{(\sqrt{\pi}R)^d},
\end{equation}
an equation which has been obtained before too \cite{Shirts2006,Uline2008}.

The direct marginalization \cite{Song1997} of the joint PDF of all velocities,
Eq.~\eqref{eq:uniformprobability2}, leads to the PDF $f_{v_{i\alpha}}(x)$ of
velocity components, a result obtained integrating over all $i$ except one
and over all $\alpha$ except one. This is the quantity discussed by Maxwell
\cite{Maxwell1860a}, and its derivation for any $N$ is one of the main results
in this paper. It turns out that the distribution of the velocity components
is a transformed beta with argument $1/2+x/(2R)$ and equal exponents
$a = b = (dN-1)/2$,
\begin{align}
\label{eq:fvia}
f_{v_{i\alpha}}(x) =& \frac{1}{\mathrm{B}\big((dN-1)/2,(dN-1)/2\big)}
\nonumber \\
& \times \left[ \left(\frac{1}{2}+\frac{x}{2R}\right)
\left(\frac{1}{2}-\frac{x}{2R}\right) \right]^\frac{dN-3}{2}
\frac{\mathbf{1}_{[-R,R]}(x)}{2R}
\nonumber\\
=& \frac{\Gamma(dN-1)}{\Gamma^2\big((dN-1)/2\big)}
\left(1-\frac{x^2}{R^2}\right)^\frac{dN-3}{2}
\frac{\mathbf{1}_{[-R,R]}(x)}{2^{dN-2}R}
\nonumber \\
=& f_{v_{i\alpha}}^\beta \left(\frac{1}{2} + \frac{x}{2R};
\frac{dN-1}{2},\frac{dN-1}{2}\right) \nonumber \\
& \times \frac{\mathrm{d}}{\mathrm{d}x}\left(\frac{1}{2}+\frac{x}{2R}\right).
\end{align}
In the thermodynamic limit Eq.~(\ref{eq:fvia}) converges to a Gaussian with
average $\mu = 0$ and variance $\sigma^2 = R^2/(dN) = d\bar{E}/(2m)$, i.e.\
the familiar Maxwell-Boltzmann distribution
\begin{equation}
\label{eq:normal}
f_{v_{i\alpha}}(x) = \sqrt{\frac{m}{d\pi\bar{E}}}
\exp\left(-\frac{mx^2}{d\bar{E}}\right).
\end{equation}
This is again related to a gamma distribution, since the positive half of a
Gaussian can be expressed as
\begin{equation}
\frac{2}{\sqrt{2\pi}\sigma}\exp\left(-\frac{x^2}{2\sigma^2}\right)
\mathbf{1}_{[0,\infty)}(x) = f_X^\gamma\left(\frac{x^2}{2};\frac{1}{2},
\sigma^2\right) \frac{\mathrm{d}}{\mathrm{d}x} \frac{x^2}{2}.
\end{equation}

In summary, all the known results for the relevant distributions of the $NVE$
ensemble can be obtained observing that the normalized individual particle
energies $\varepsilon_i = E_i/E$ follow a symmetric multivariate Dirichlet
distribution with parameter $a = 1/2$ given by Eq.~\eqref{eq:dirichlet1}.
This is a direct consequence of the uniform-distribution assumption in
Eq.~\eqref{eq:uniformprobability1} via a simple change of variables. Only for
the velocity components it is necessary to marginalize the uniform distribution
directly on the surface of the hypersphere and not on the simplex. Maxwell's
\emph{Ansatz} is vindicated by the fact that, in the thermodynamic limit,
a normal distribution for velocity components is recovered, as well as their
independence. Finally, in the $NVE\mathbf{PG}$ ensemble the constraint given
by Eq.~\eqref{eq:momentum} leads to similar distributions with different
parameter values for the relevant quantities introduced above. This will
become clearer in the following.

\section{Molecular dynamics simulations}
\label{sec:md}

In molecular dynamics (MD) with continuous potentials, the equations of motion
are integrated numerically using a constant time step; this approach is called 
time-driven. The larger the forces, the smaller the time step necessary to
ensure energy conservation. With step potentials there are no forces acting
on a distance, only impulsive ones at the exact time of impact. Therefore an
event-driven approach is more appropriate: rather than until a fixed time step,
the system is propagated until either the next collision or the next boundary
crossing \cite{Alder1957,Alder1959,Allen1989,Rapaport2004}.

The collision time $t_{ij}$ between two particles $i,j$ can be calculated from
the mutual distance $\mathbf{r}_{ij} = \mathbf{r}_i - \mathbf{r}_j$ and the
relative velocity $\mathbf{v}_{ij} = \mathbf{v}_i - \mathbf{v}_j$. If $b_{ij}
= \mathbf{v}_{ij} \cdot \mathbf{r}_{ij} > 0$ the particles are moving away from
each other and will not collide. Otherwise impact may happen at time $t_{ij}$
when their distance becomes equal to the sum of their radii, i.e.,
$\| \mathbf{r}_{ij} + t_{ij} \mathbf{v}_{ij} \| = \sigma$. 
This is a second order problem with solutions
\begin{equation}
t_{ij}^\pm = \frac{-b_{ij} \pm \sqrt{b_{ij}^2 - v_{ij}^2 (r_{ij}^2 -
\sigma^2)}}{v_{ij}^2}.
\end{equation}
If the solutions are complex, no collision occurs. If the solutions are real,
the smaller one, $t_{ij}^-$, corresponds to when the particles first meet,
while the larger one, $t_{ij}^+$, to when they leave each other assuming they
are allowed to interpenetrate. A negative collision time means that the event
took place in the past. Because of the condition $b_{ij} < 0$, at least
$t_{ij}^+ > 0$. If $t_{ij}^- < 0$ the particles overlap, which indicates an
error. So the collision time is given by $t_{ij}^-$, provided it is a positive
real number.

For a system of $N$ hard spheres, at impact, assuming an elastic collision,
the total kinetic energy $E$ and the total linear momentum $\mathbf{P}$ 
are conserved (usually one sets $\mathbf{P = 0}$ at the beginning of the
simulation by subtracting $\frac{1}{N}\sum_{i=1}^N \mathbf{v}_i$ from each
$\mathbf{v}_i$). Assuming smooth surfaces, the impulse acts along the line
of centers of the collision partners $i$ and $j$ given by the unit vector
$\hat{\mathbf{r}}_{ij} = \mathbf{r}_{ij}/r_{ij}$; with equal masses,
$\mathbf{v}_i$ changes to $\mathbf{v}_i + \Delta \mathbf{v}_i$ and
$\mathbf{v}_j$ changes to $\mathbf{v}_j - \Delta \mathbf{v}_i$ with
\begin{equation}
\label{eq:velupdate}
\Delta \mathbf{v}_i = -\frac{b_{ij} \mathbf{r}_{ij}}{\sigma^2}
= - (\mathbf{v}_{ij} \cdot \hat{\mathbf{r}}_{ij}) \, \hat{\mathbf{r}}_{ij}
= - \mathbf{v}_{ij}^\parallel,
\end{equation}
where $\mathbf{r}_{ij}$ and $\mathbf{v}_{ij}$ are evaluated at the instant of
collision, and thus $r_{ij} = \sigma$.

A few computational details for our event-driven molecular dynamics simulation
of hard spheres are given in the appendix.

\section{Monte Carlo simulations}
\label{sec:mc}

Except for especially ordered initial conditions, interparticle collisions
computed by MD as explained in Sec.~\ref{sec:md} have mutual distance versors
at collision $\hat{\mathbf{r}}_{ij}$ uniformly distributed on a unit half
sphere in $d$ dimensions such that, given relative velocities
$\mathbf{v}_{ij}$, the scalar product $\mathbf{v}_{ij}\cdot
\hat{\mathbf{r}}_{ij}$ is negative. Therefore the same distributions of
velocities, and thus of derived quantities like energies, as in MD with
periodic boundaries can be obtained by Monte Carlo (MC): after initializing the
velocities of all hard spheres, the MC cycles consist in selecting a pair $ij$
and a random versor $\hat{\mathbf{r}}_{ij}$ such that $\mathbf{v}_{ij} \cdot
\hat{\mathbf{r}}_{ij} < 0$, and then in updating the velocities according to
Eq.~\eqref{eq:velupdate}. Hard reflecting walls can be included in the MC
scheme by selecting with a certain frequency a sphere $i$ and inverting one of
its velocity components $v_{i\alpha}$. This scheme gives a useful insight into
the mechanism of energy and momentum transfer. It is much easier to code and
faster to run than MD, especially for large numbers of particles $N$, because
no event list management is necessary: on the same computer used for the MD
benchmarks shown in Fig.~\ref{fig:benchmark_md}, the CPU time for $10^5$ MC
collisions with $N = 10\,000$ spheres is 0.3\,s.

For a given initial state, i.e.\ a set of particle velocities, the MC dynamics
defined above provides the realization of a Markov chain with a symmetric
transition kernel, meaning that $P(\mathbf{v}'|\mathbf{v}) = P(\mathbf{v}|
\mathbf{v}')$, where $\mathbf{v}$ is the old velocity vector before the
transition and $\mathbf{v}'$ is the new velocity vector after the transition.
This Markov chain is homogeneous, as the transition probability does not
depend on the time step. Invoking detailed balance, $P(\mathbf{v}'|\mathbf{v})
P(\mathbf{v}) = P(\mathbf{v}|\mathbf{v}') P(\mathbf{v}')$, the symmetry of
the transition kernel implies that the stationary distribution of this chain
is uniform over the set of accessible states. If this set coincides with the
surface of the velocity hypersphere, then the Markov chain is ergodic and
one can hope to prove that the uniform distribution over the hypershere is
also the equilibrium distribution for the Markov chain; see Sigurgeirsson
\cite[Chapter 5]{Sigurgeirsson2002} for the discussion of a related problem,
and Meyn and Tweedie \cite{Meyn1993} for general methods. The results of
MC simulations described below corroborate this conjecture and the algorithm
outlined above is indeed an effective way of sampling the uniform distribution
on the surface of a hypersphere.

\section{Numerical results}
\label{sec:results}

\begin{figure*}[htbp]
\includegraphics[width=\columnwidth]{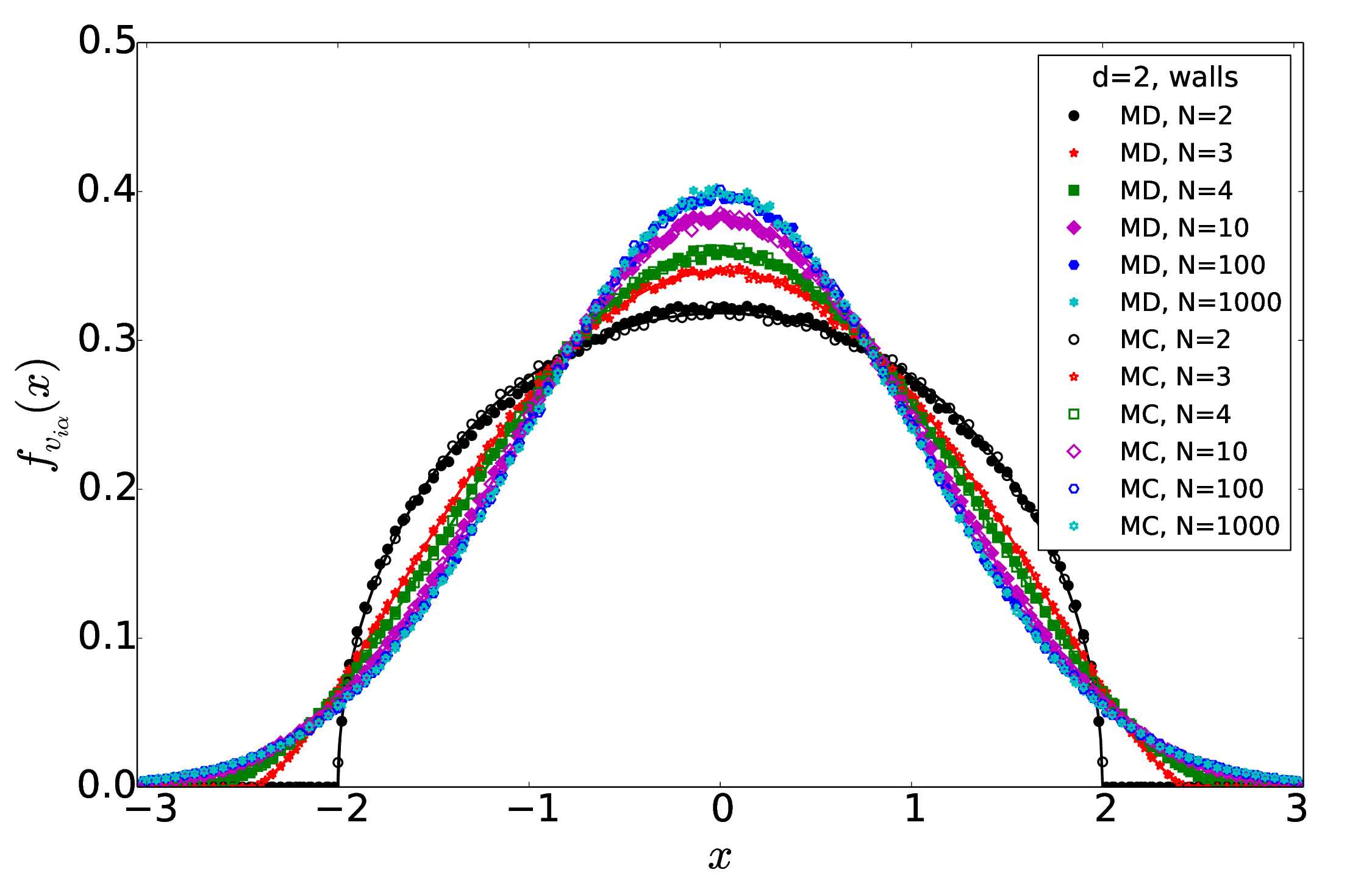}
\includegraphics[width=\columnwidth]{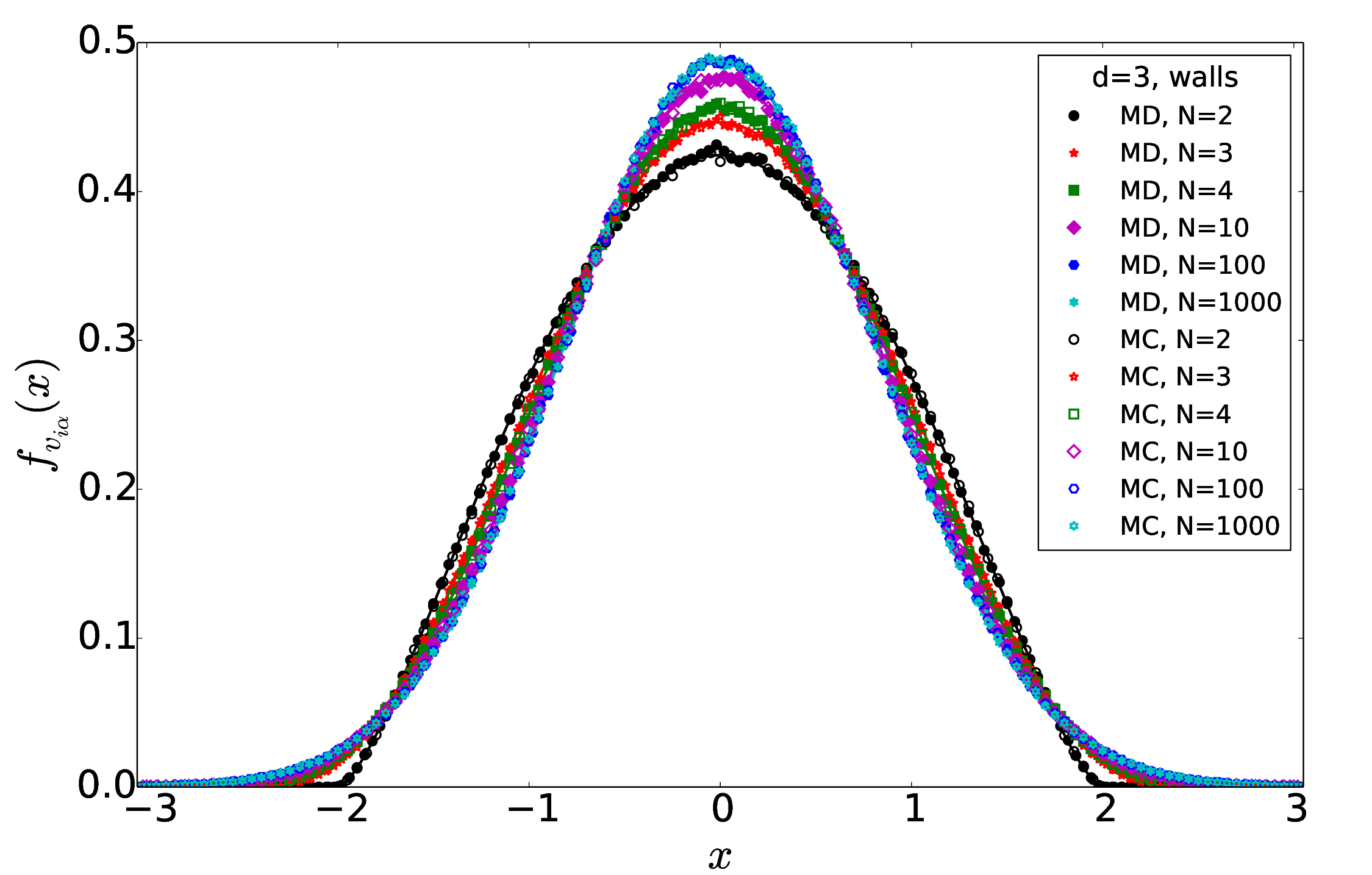}
\includegraphics[width=\columnwidth]{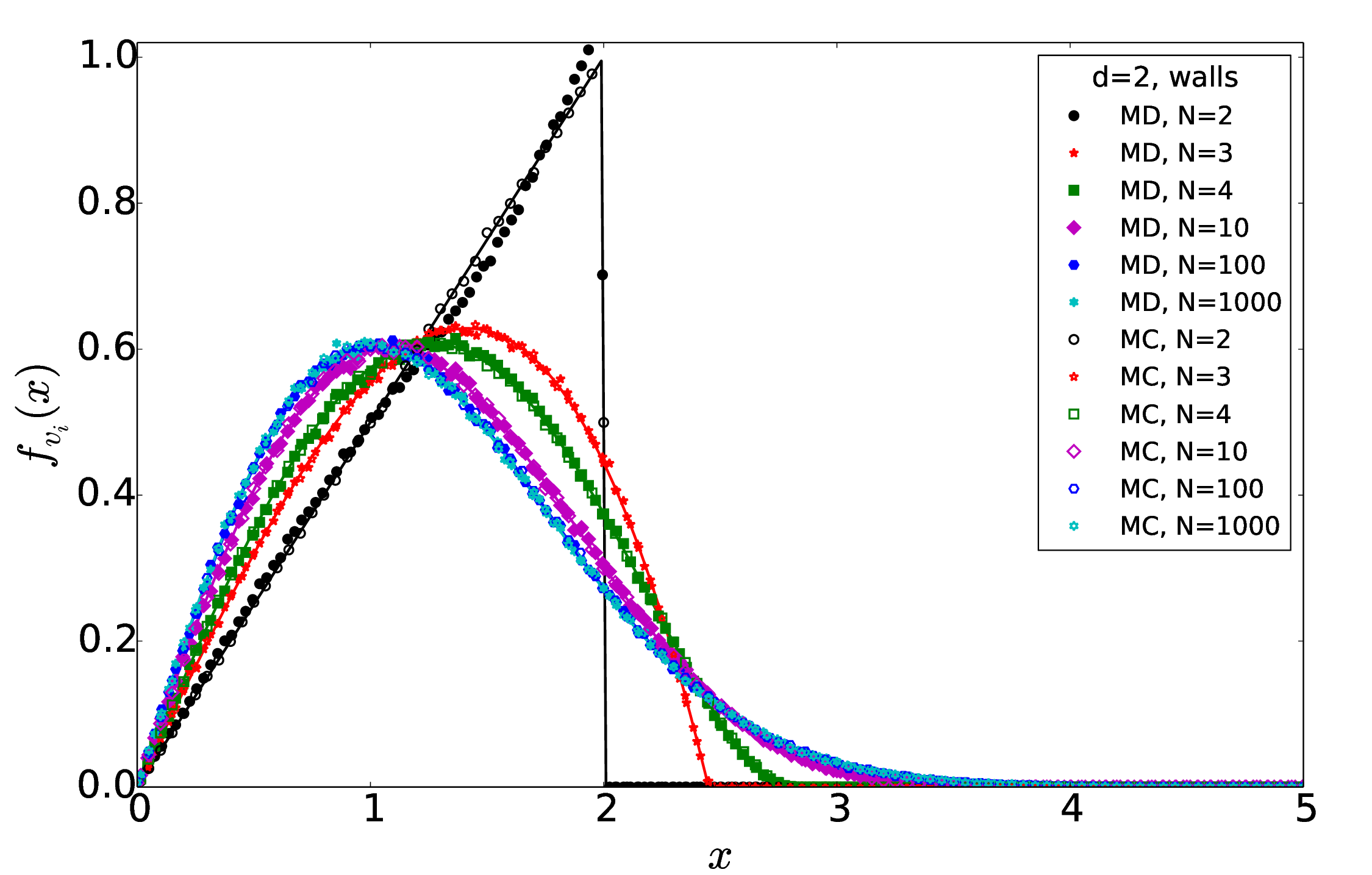}
\includegraphics[width=\columnwidth]{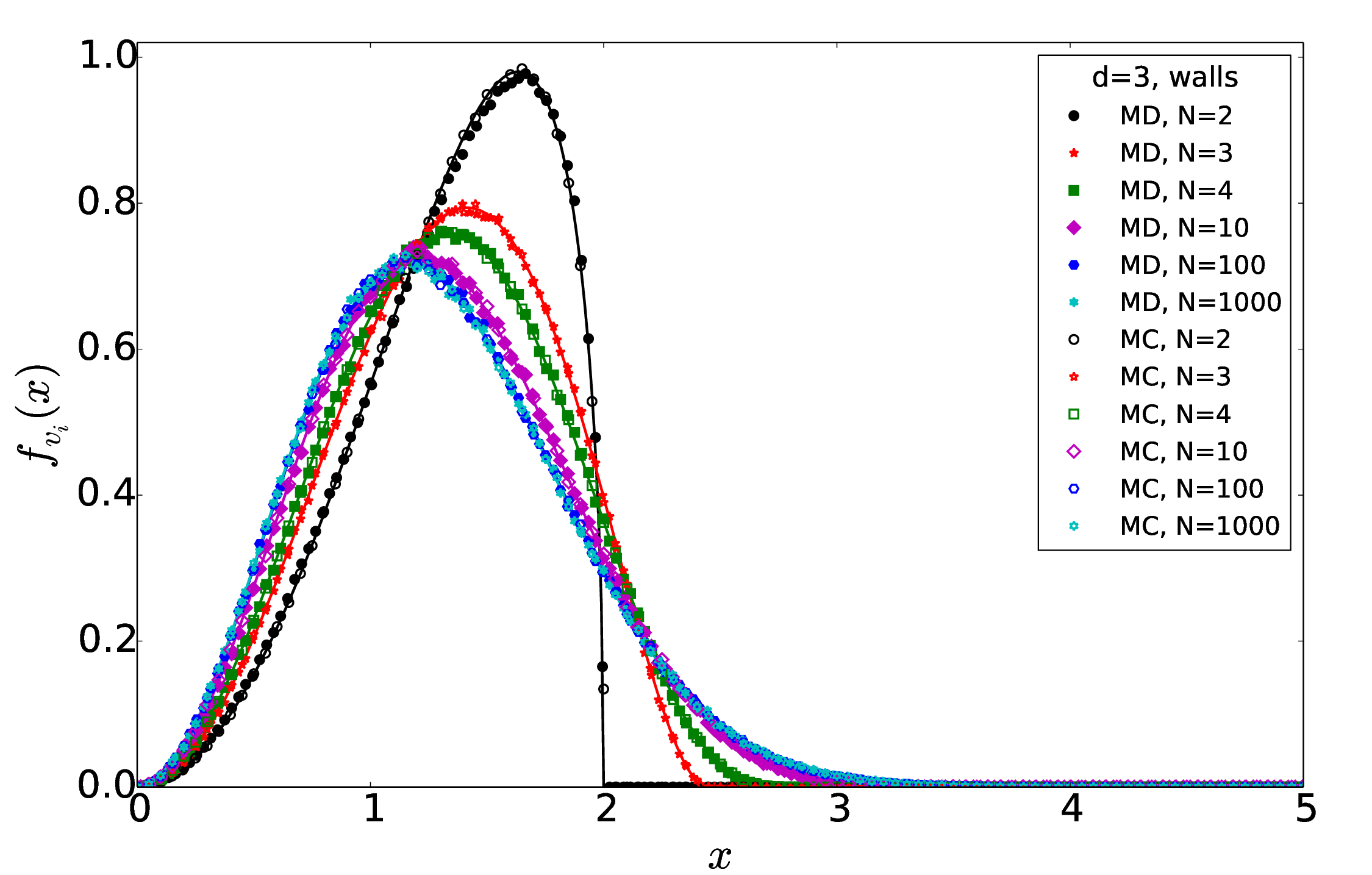}
\includegraphics[width=\columnwidth]{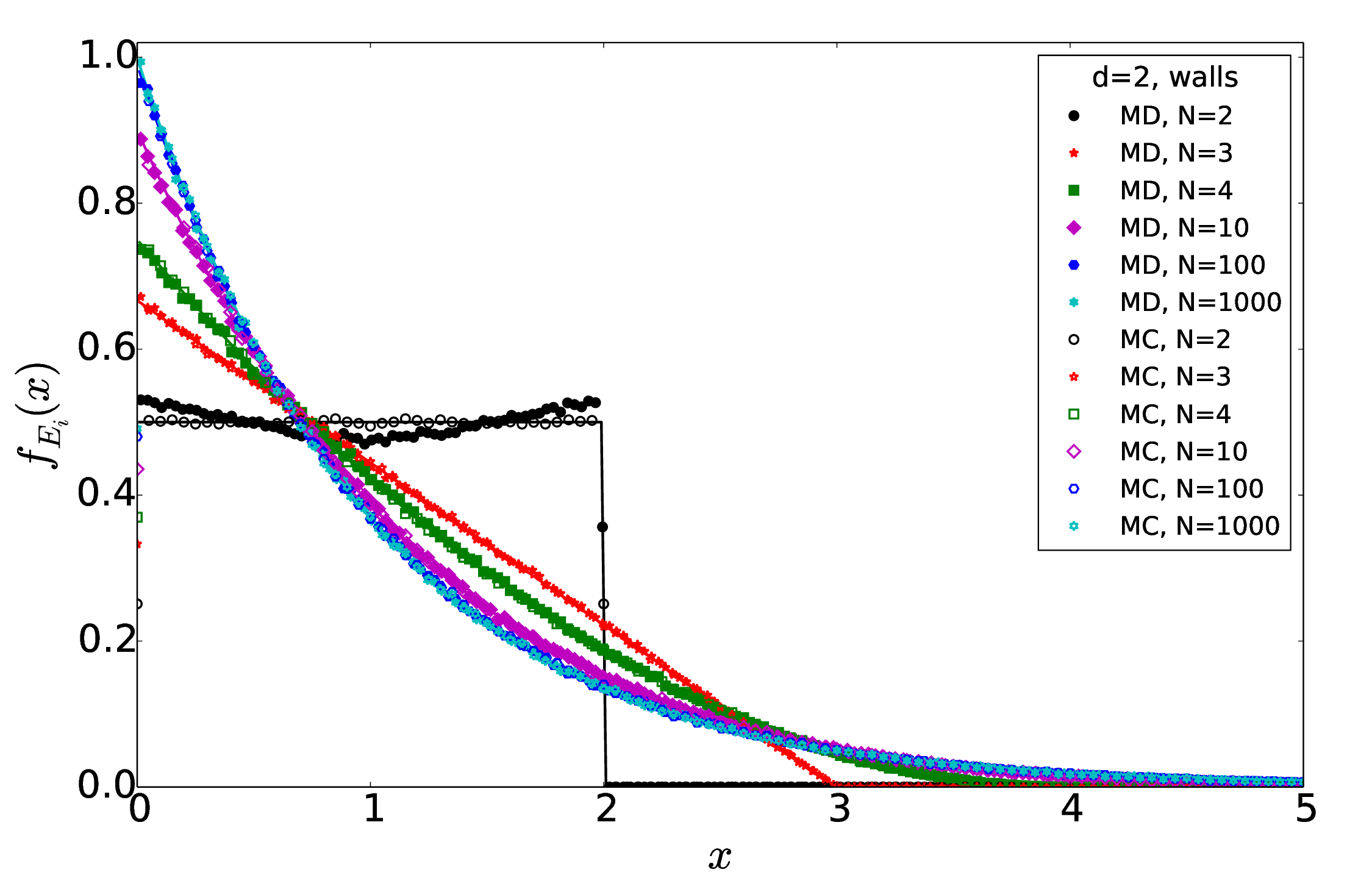}
\includegraphics[width=\columnwidth]{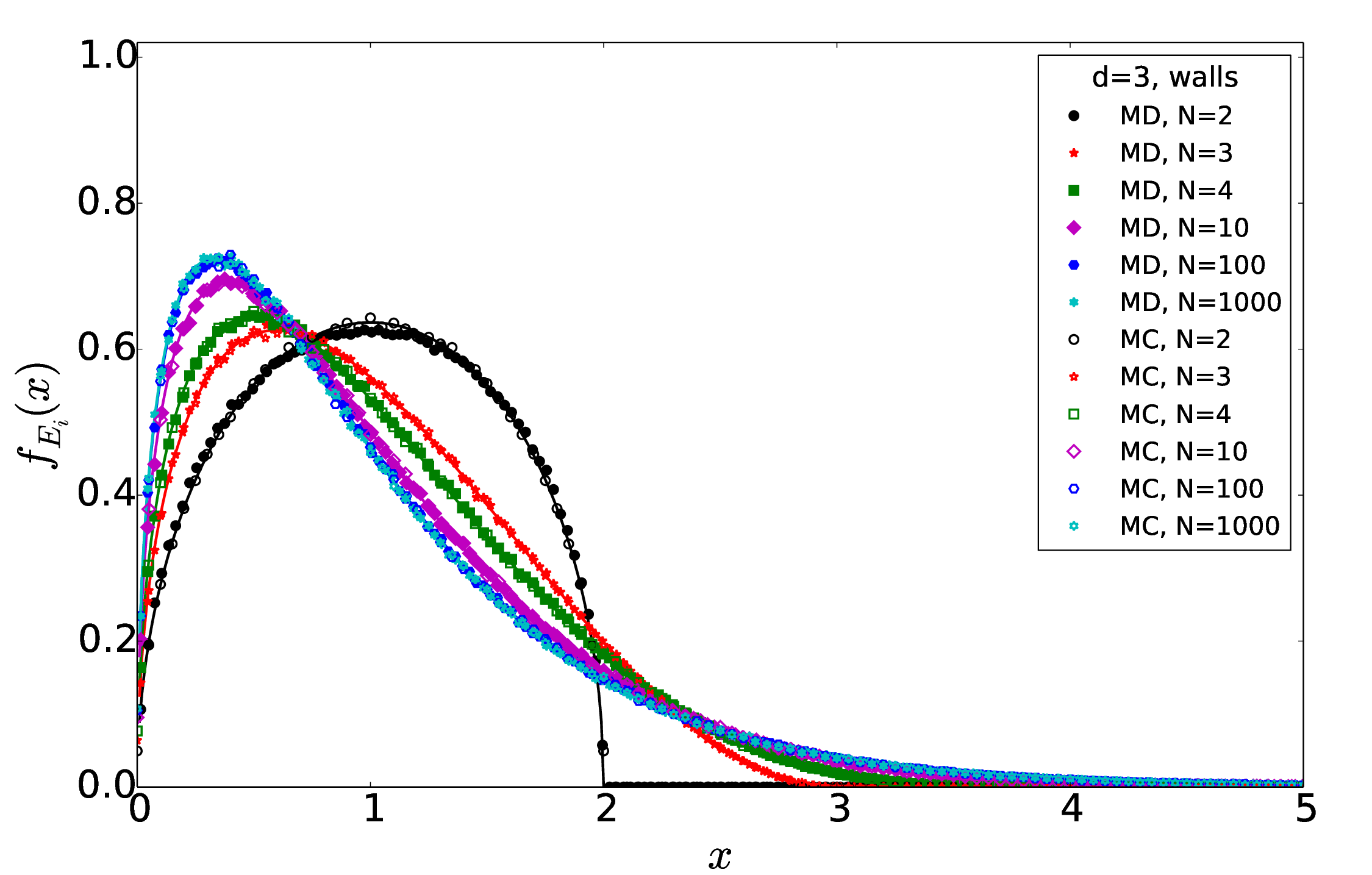}
\caption{\label{fig:results_walls}
(Color online) Probability density functions $f_{v_{i\alpha}}$ of the velocity
components (top), $f_{v_i}$ of the velocity modulus (middle), and $f_{E_i}$ of
the energy (bottom) for $d = 2$ (left) and $d = 3$ (right) with $\bar{E} = 1$
in the microcanonical ensemble (constant $NVE$, hard reflecting walls): theory
(lines), MD (full symbols), and MC (empty symbols).}
\end{figure*}

\begin{figure*}[htbp]
\includegraphics[width=\columnwidth]{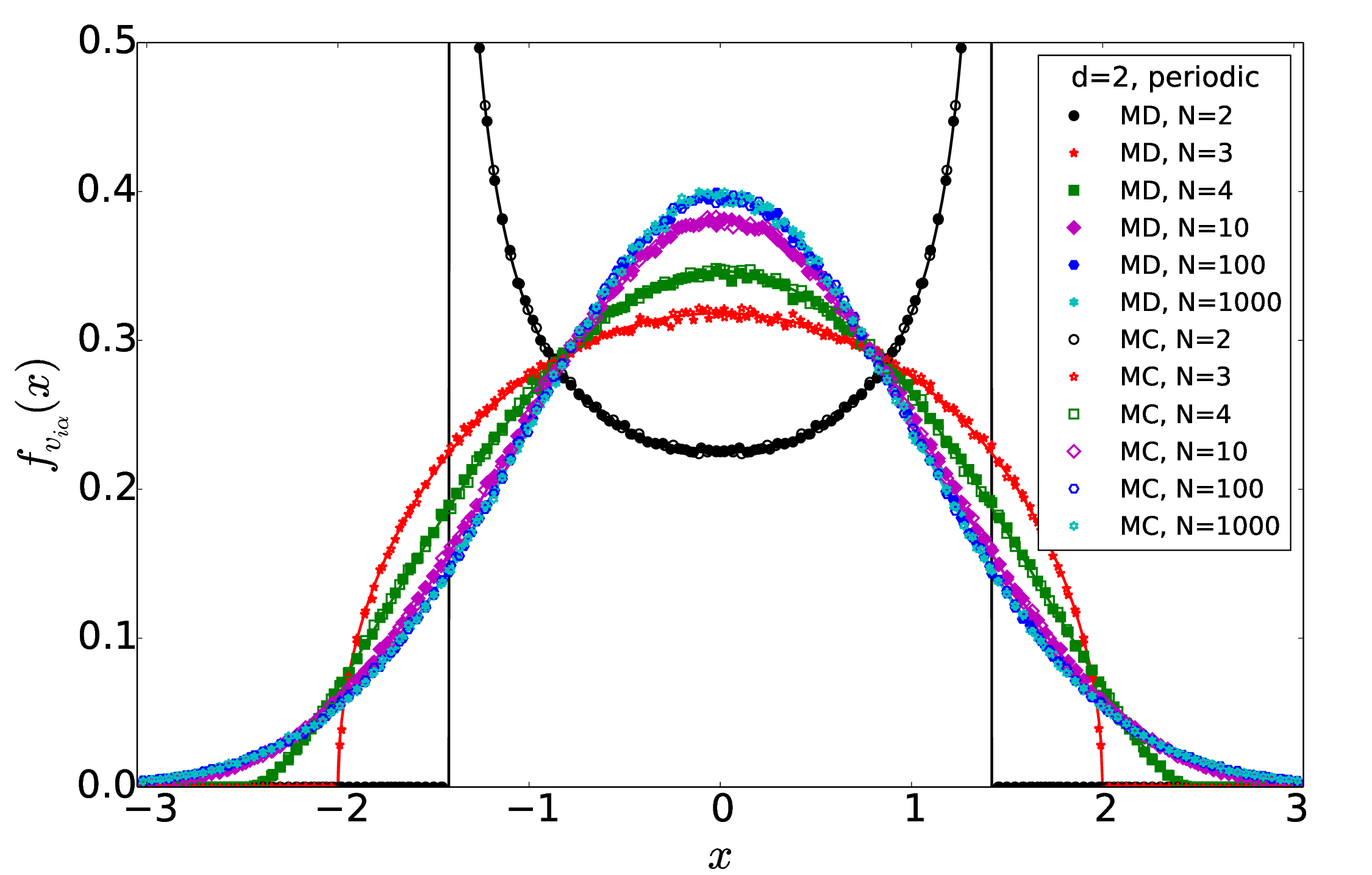}
\includegraphics[width=\columnwidth]{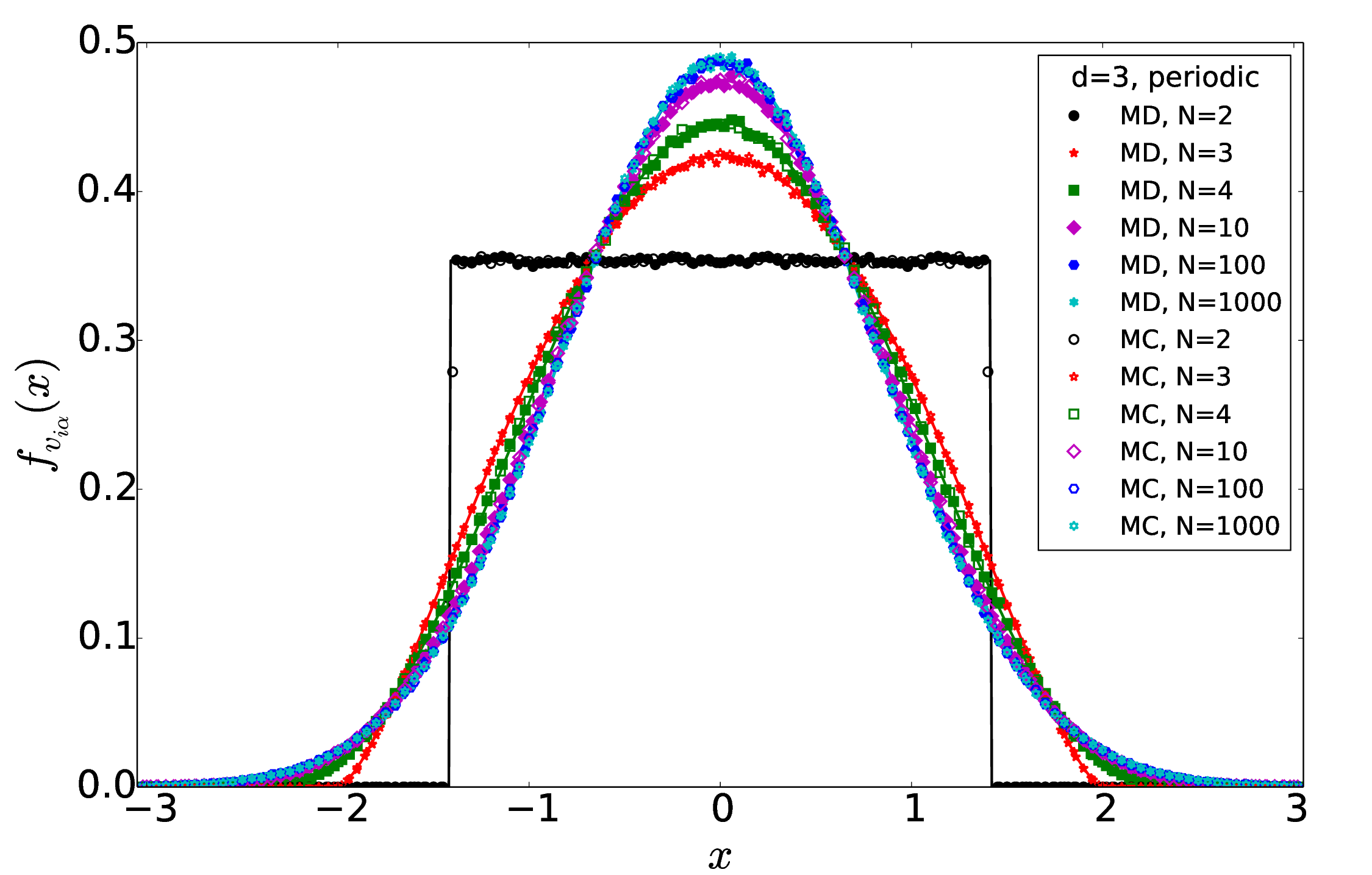}
\includegraphics[width=\columnwidth]{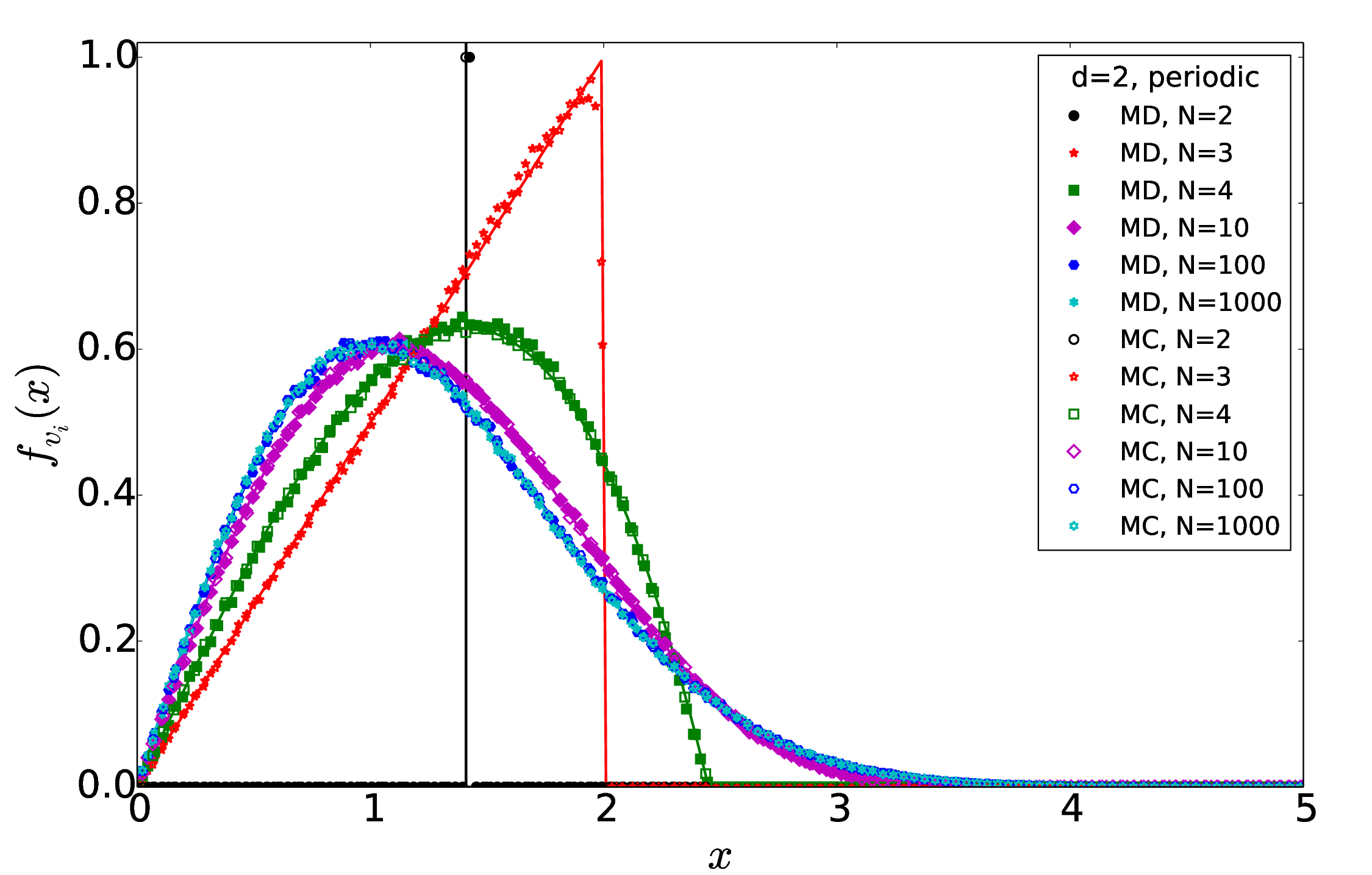}
\includegraphics[width=\columnwidth]{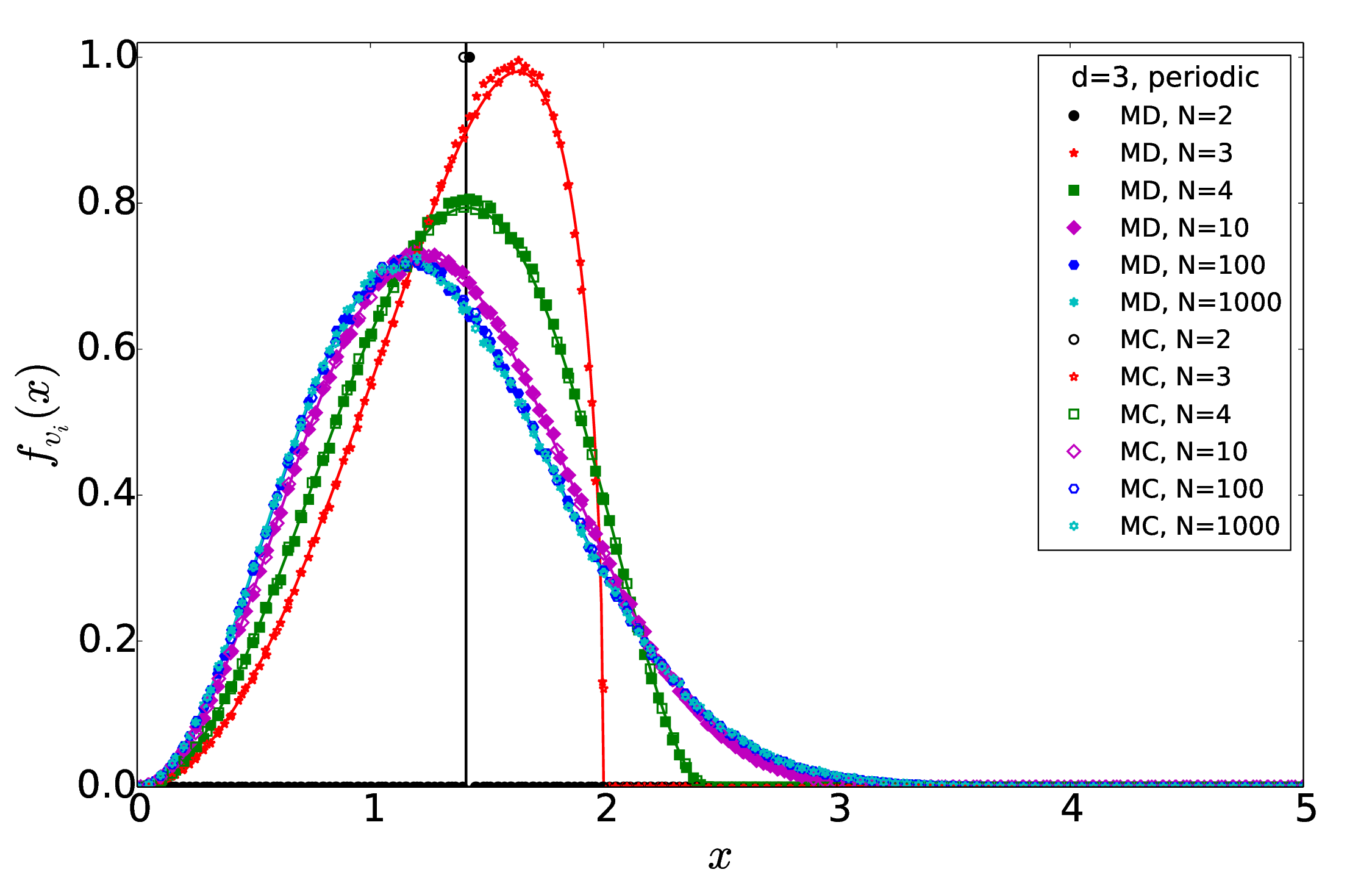}
\includegraphics[width=\columnwidth]{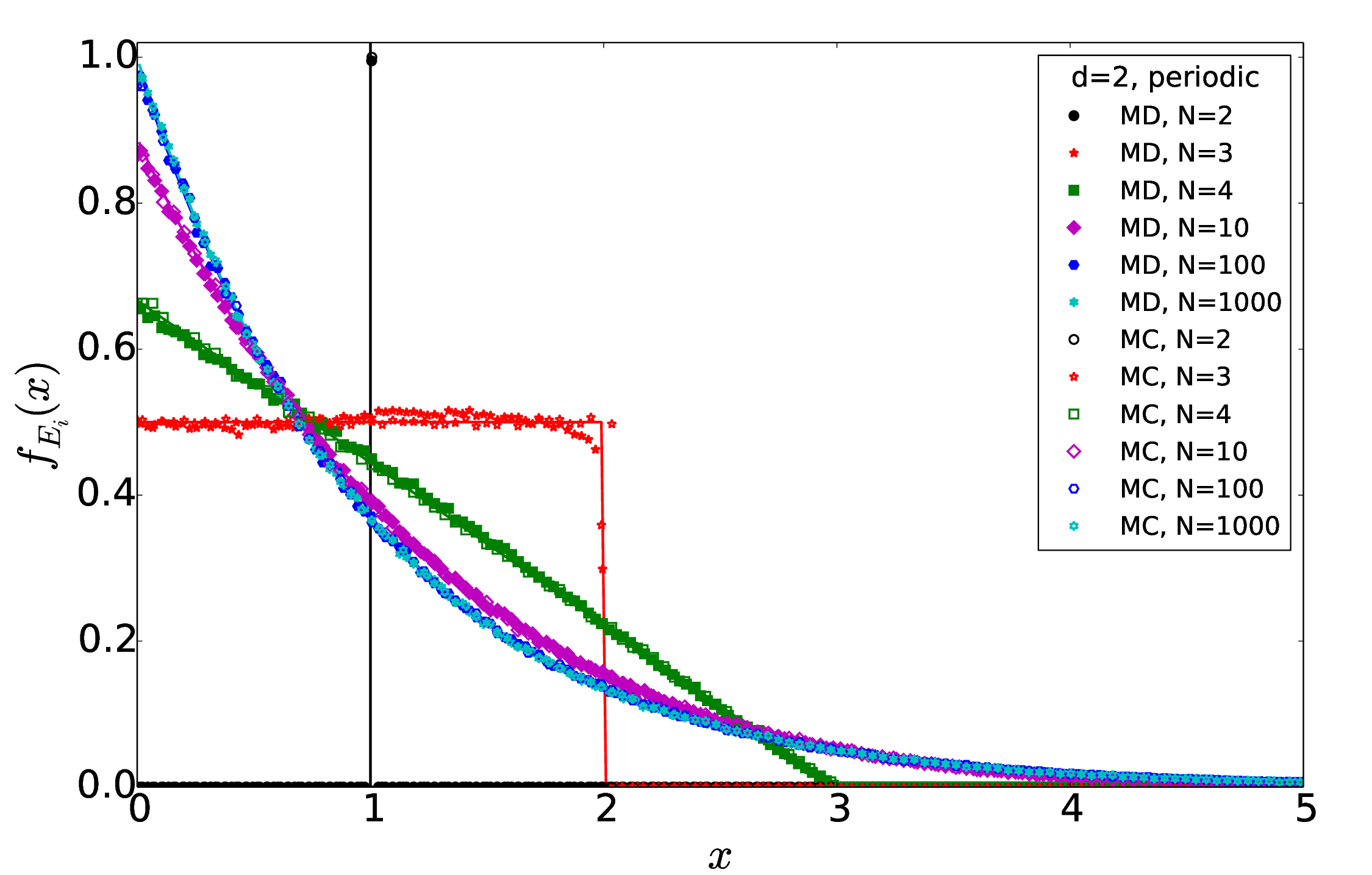}
\includegraphics[width=\columnwidth]{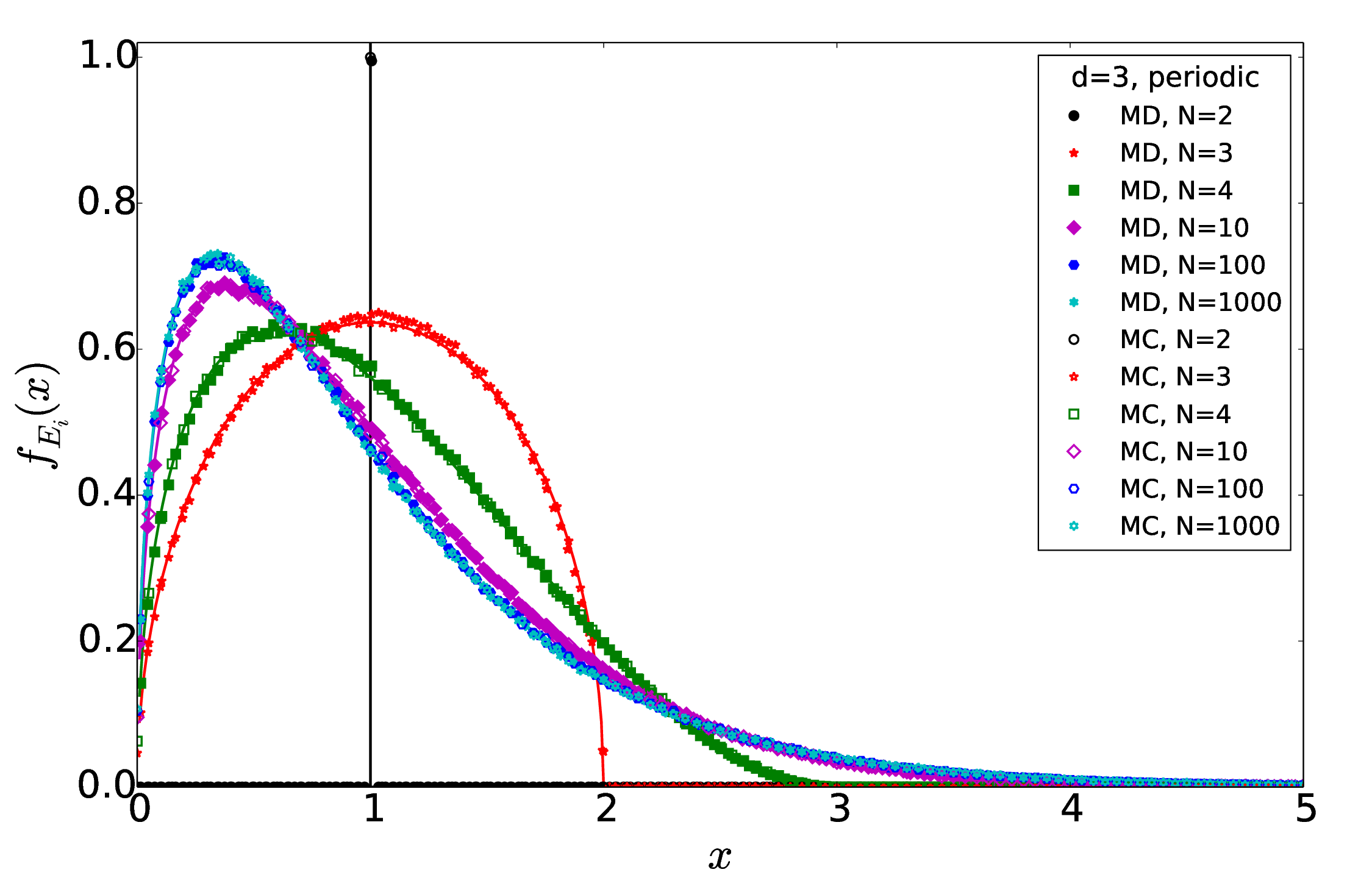}
\caption{\label{fig:results_periodic}
(Color online) Probability density functions $f_{v_{i\alpha}}$ of the velocity
components (top), $f_{v_i}$ of the velocity modulus (middle), and $f_{E_i}$ of
the energy (bottom) for $d = 2$ (left) and $d = 3$ (right) with $\bar{E} = 1$
in the molecular dynamics ensemble (constant $NVE\mathbf{PG}$, periodic
boundary conditions): theory (lines), MD (full symbols), and MC (empty
symbols). Delta functions are made visible by a vertical line for the theory
and by rescaling down to 1 the data point that would otherwise be out of
scale.}
\end{figure*}

Probability density functions $f_{v_{i\alpha}}$ of the velocity components,
$f_{v_i}$ of the velocity modulus, and $f_{E_i}$ of the energy for $N$ = 2, 3,
4, 10, 100, 1000 hard disks ($d = 2$) and hard spheres ($d = 3$) from theory
(Sec.~\ref{sec:theory}) as well as from MD (Sec.~\ref{sec:md}) and MC
simulations (Sec.~\ref{sec:mc}) are shown in Fig.~\ref{fig:results_walls}
for a microcanonical ensemble (constant $NVE$, hard reflecting walls)
and in Fig.~\ref{fig:results_periodic} for a molecular dynamics ensemble
(constant $NVE\mathbf{PG}$, periodic boundary conditions). Simulations with
$N = 10\,000$ were done too, but are not shown because the distributions
overlap perfectly with those where $N = 1000$. In reduced units \cite{Allen1989}
the particle mass is $m = 1$, the particle diameter is $\sigma = 1$, the energy
per particle is $\bar{E} = 1$, the Boltzmann constant is $k_\mathrm{B} = 1$,
and the number density is $\rho = 2/3^d$. The density appears only in MD, and
the results are largely independent of this parameter, as long as it is not too
large (in this case the particles cannot move freely) or too small (in this
case the particles hardly ever collide). The initial total momentum is
$\mathbf{P} = \boldsymbol{0}$, except for $N = 2$ with hard reflecting walls,
and the initial position of the center of mass is in the origin. The numerical
simulations were equilibrated over $5 \times 10^5$ collisions and sampled over
$10^6$ collisions.

We did not do the case with $d = 1$ because then with equal masses
Eq.~(\ref{eq:velupdate}) becomes $\Delta \mathbf{v}_i =
-\mathbf{v}_{ij}^\parallel = -\mathbf{v}_{ij}$; thus after a collision
$\mathbf{v}_i$ becomes $\mathbf{v}_j$ and vice versa, and with particles just
exchanging their velocities, the velocity distribution does not change with
time, making equilibration impossible. For this reason, one-dimensional models
need systems of molecules with different masses \cite{Boozer2011}.

The agreement between theory, MD and MC is excellent, with little systematic
deviations only for MD at the smallest values of $N$. To check that these
deviations are genuine, we tried with different starting configurations and
densities, and with both MD programs discussed in Sec.~\ref{sec:md}, obtaining
identical results. For further details on our MD simulations see Gabriel
\cite[Chapter 3]{Gabriel2010}.

For $d = 2$ and $N = 2$ in the $NVE\mathbf{PG}$ ensemble $f_{v_{i\alpha}}(x)$
is the arcsine PDF
\begin{equation}
\label{eq:arcsine}
f_{v_{i\alpha}}(x) = \frac{1}{\pi\sqrt{R^2 - x^2}},
\end{equation}
which is bimodal. The name is due to its cumulative distribution function
\begin{equation}
\label{eq:arcsine_cumul}
F_{v_{i\alpha}}(x) = \frac{1}{\pi} \arcsin\frac{x}{R} + \frac{1}{2}.
\end{equation}
For $d = 2$ and $N = 2$ in the $NVE$ ensemble and for $d = 2$ and $N = 3$ in
the $NVE\mathbf{PG}$ ensemble, $f_{v_{i\alpha}}(x)$ is the Wigner semicircle
PDF \cite{Politi2010}
\begin{equation}
\label{eq:semicircle}
f_{v_{i\alpha}}(x) = \frac{2}{\pi R^2} \sqrt{R^2 - x^2}.
\end{equation}
Its cumulative distribution function
again contains an arcsine:
\begin{equation}
\label{eq:semicircle_cumul}
F_{v_{i\alpha}}(x) = \frac{x}{\pi R^2} \sqrt{R^2 - x^2}
+ \frac{1}{\pi} \arcsin\frac{x}{R} + \frac{1}{2}.
\end{equation}
For the same systems, $f_{E_i}(x)$ is the uniform distribution on $[0,R]$.
For $d = 3$ and $N = 2$ in the $NVE\mathbf{PG}$ ensemble, $f_{v_{i\alpha}}(x)$
is the uniform distribution on $[-\sqrt{R},\sqrt{R}]$.
All distributions are given by Eqs.~(\ref{eq:fEi}), (\ref{eq:fvi}) and
(\ref{eq:fvia}) inserting the appropriate values of $d$, $N$ and $E = N\bar{E}$
or $R = \sqrt{2E/m}$ for the $NVE$ ensemble, while for the $NVE\mathbf{PG}$
ensemble $N$ must be substituted by $N-1$ because of the additional constraint
on the linear momentum and thus on the center of mass.

To quantify the visual impression, in Tab.~\ref{tab:ks} we show
Kolmogorov-Smirnov goodness-of-fit tests \cite{Kolmogorov1933,Smirnov1939,
Massey1951} comparing Eq.~(\ref{eq:fvia}) with the empirical cumulative
distribution function of the MC velocity components in the $NVE\mathbf{PG}$
ensemble for $d = 2$ and $N$ = 2, 3, 10, 100, 1000, 10\,000. In all cases the
null hypothesis of data distributed according to the model equation cannot be
rejected at the 5\% significance level.
The empirical density $f_{v_{i\alpha}}(x)$ is well approximated by a normal law
for $N \geq 1000$ hard disks, as shown in Tab.~\ref{tab:gaussian}, where the
results of two non-parametric tests for normality, Lilliefors
\cite{Lilliefors1967} and Jarque and Bera \cite{Jarque1980,Jarque1981}, are
presented for the MC velocity components of the systems with periodic boundary
conditions, $d = 2$ and $N = 10, 100, 1000, 10\,000$; these tests were made
with the \textsc{Matlab} functions lillietest and jbtest. 

\begin{table}[htb]
\begin{ruledtabular}
\begin{tabular}{ccc}
    $N$ & Kolmogorov-Smirnov & $p_\mathrm{KS}$ \\
\hline
      2 & $4.1 \times 10^{-4}$ & 0.99 \\
      3 & $1.0 \times 10^{-3}$ & 0.24 \\
     10 & $1.3 \times 10^{-3}$ & 0.07 \\
    100 & $7.1 \times 10^{-4}$ & 0.69 \\
   1000 & $6.2 \times 10^{-4}$ & 0.84 \\
10\,000 & $1.1 \times 10^{-3}$ & 0.15
\end{tabular}
\end{ruledtabular}
\caption{\label{tab:ks}
Comparison between the empirical probability density functions of the velocity
components from MC with periodic boundary conditions and $d = 2$ by means of the
Kolmogorov-Smirnov (KS) test. In each case the sample size is $2 \times 10^6$.
At the 5\% significance level the critical value is $9.6 \times 10^{-4}$.
The null hypothesis of equally distributed data can never be rejected.}
\end{table}

\begin{table}[h]
\begin{ruledtabular}
\begin{tabular}{ccccc}
 $N$ & Lilliefors & $p_\mathrm{L}$ & Jarque-Bera & $p_\mathrm{JB}$ \\
\hline
     10 &              0.008* & $< 10^{-3}$ & $7.4\times 10^3$* & $< 10^{-3}$\\
    100 & $7.36\times 10^{-4}$* &      0.01 &             29.6* & $< 10^{-3}$\\
   1000 & $4.79\times 10^{-4}$  &      0.35 &             5.70  &        0.06\\
10\,000 & $4.37\times 10^{-4}$  &      0.50 &             2.36  &        0.31
\end{tabular}
\end{ruledtabular}
\caption{\label{tab:gaussian}
Results of two non-parametric normality tests for the empirical probability
density function of the velocity components from MC with periodic boundary
conditions when $d = 2$: Lilliefors (L) and Jarque and Bera (JB). In each case
the sample size is $2 \times 10^6$. At the 5\% significance level the critical
value is $6.43 \times 10^{-4}$ for the L test and 5.99 for the JB test. The
star indicates that the null hypothesis of normally distributed data can be
rejected.}
\end{table}

\section{Discussion and conclusions}
\label{sec:conclusions}

To summarize what we have done, in a system of $N$ hard balls in a
$d$-dimensional volume $V$ the velocity components, the velocity modulus and
the energies of the spheres or disks are well reproduced by transformed beta
distributions with different arguments and shape parameters depending on $N$,
$d$, the total energy $E$, and the boundary conditions; in the thermodynamic
limit these distributions converge to transformed gamma distributions with
different arguments and shape or scale parameters, corresponding respectively
to the Gaussian, i.e.\, Maxwell-Boltzmann, the Maxwell, and the Boltzmann or
Boltzmann-Gibbs distribution. We showed this theoretically using Khinchin's
\emph{Ansatz}, and performed statistical goodness-of-fit tests on systematic
MD and MC computer simulations of an increasing number $N$ of hard disks or
spheres starting from 2 in the microcanonical ensemble (constant $NVE$,
hard reflecting walls) and in the molecular dynamics ensemble (constant
$NVE\mathbf{PG}$, periodic boundary conditions). The MC simulations are a
simple stochastic model based on a generalization of Eq.~(\ref{eq:randomwalk})
able to reproduce the same empirical equilibrium distribution for the random
variables $v_{i\alpha}$, $v_i$ and $E_i$ as obtained deterministically with
canonic dynamics by MD simulations. While the MC results overlap perfectly
with theory, there is a slight disagreement of the MD results for the smallest
values of $N$. A definitive explanation of this will require further
investigation. However, one can recall that our MC scheme acts only
on velocities independently of positions and the volume $V$, with the
interparticle versor at collision sampled uniformly on a unit half sphere
in $d$ dimensions. On the contrary, due to the canonical dynamics, in MD
velocities and positions are mutually dependent; therefore geometrical
constraints in the smallest systems may lead to a non perfectly uniform
sampling of the state space. Moreover, in the MC systems with walls the
correlation among velocities at impact is below 0.001 for every value of $N$
(random variables uniformly distributed on the hypersphere are uncorrelated
even if dependent), so that the assumptions of molecular chaos are always
fulfilled, whereas a computer simulation of a one-dimensional system has shown
that in MD this happens only with growing $N$ \cite{Boozer2011}.

We presented comprehensively both analytical derivations with a new approach
and numerical checks, the latter both by MD and MC with a systematic
investigation of parameter values and boundary conditions; we obtained also the
PDF of velocity components, Eq.~(\ref{eq:fvia}), which was not possible with
previous approaches; we realized that all these distributions are variants
of the beta or the gamma distribution, and can be derived from the Dirichlet
distribution; we pointed out that for values of $N$ as low as 2 or 3 the shapes
of these distributions can be quite different from those in the thermodynamic
limit: in particular, they can become uniform or even bimodal; last, we
discussed the slight deviations of the MD results from theory.

The significance of our investigations goes beyond the foundations of
statistical mechanics: few-body systems and microclusters are of current
practical interest in applied fields such as nanotechnology and biophysics,
and there is an increasing effort to understand the statistical mechanics of
such systems, which departs from the traditional approach in the thermodynamic
limit \cite{Niiyama2009}. Another application of our results may be in
the theory of thermostats or heat baths with a finite thermal capacity
\cite{Campisi2009}.

The MD simulations presented above corroborate Boltzmann's ergodic hypothesis
\cite{Szasz1996} for both the $NVE$ and the $NVE\mathbf{PG}$ ensembles.
Sinai \cite{Sinai1967} updated this hypothesis translating it in modern
mathematical terms. One should prove that every hard-ball system on a flat
torus, after fixing its total energy, momentum, and center of mass, is fully
hyperbolic and ergodic; hyperbolic means that its Lyapunov exponent is non-zero
almost everywhere with respect to the Liouville measure. This rephrasing of
Boltzmann's hypothesis is known as the Boltzmann-Sinai ergodic hypothesis.
The proofs of ergodicity for similar systems use the so-called Chernov-Sinai
\emph{Ansatz}, namely the almost sure hyperbolicity of singular orbits
\cite{Sinai1987}. More recently, after proving the ergodiciy of hard disks
\cite{Simanyi2003}, Sim\'anyi published a proof of the Boltzmann-Sinai ergodic
hypothesis in full generality for hard-ball systems \cite{Simanyi2010b}.

MD simulations show that Khinchin's \emph{Ansatz} is justified for systems
of hard balls. We would like to stress that this is a consequence of the
microscopic dynamics and not of any a priori maximum-entropy principle.
The uniform distribution on the accessible phase-space region is indeed the
maximum-entropy distribution. Therefore, maximum-entropy methods do work
well and all the distributions in Sec.~\ref{sec:theory} could be obtained by
maximum-entropy methods: the beta and gamma distributions are actually the
maximum-entropy distributions with given first moment, and possibly some other
constraint, on a finite and a semi-infinite interval respectively. However,
this is so only because the dynamics uniformly samples the accessible
phase-space region and not the other way round. In different frameworks,
e.g.\ in biology or economics, maximum-entropy assumptions might lead to wrong
results for the equilibrium distribution of a system, if its dynamics is not
specified or carefully studied.

The distributions derived in Sec.~\ref{sec:theory} are a benchmark for random
partition models popular in econophysics. Pure exchange models often lead to
the same distributions \cite{Garibaldi2010,Patriarca2005,Patriarca2006,
Scalas2006,Garibaldi2007,Matthes2008,Chakraborti2010}.

\section*{Acknowledgments}

We are thankful to Ubaldo Garibaldi for pointing us to the Dirichlet
distribution as the multivariate generalization of the beta distribution,
and to Mauro Politi for observing that the beta distribution maximizes the
entropy on a finite interval given two constraints that include the first
moment, even if we did not use this property in the theory exposed in
Sec.~\ref{sec:theory}.
This paper was partially supported by Italian MIUR grant PRIN 2009 H8WPX5
``The growth of firms and countries: distributional properties and economic
determinants --- Finitary and non-finitary probabilistic models in economics''.
The support of the UK Economic and Social Research Council (ESRC) in funding
the Systemic Risk Centre is gratefully acknowledged (grant number ES/K002309/1).

\section*{Appendix: Computational details of the molecular dynamics simulations}

When a particle reaches a side of the unit box, periodic boundary conditions
may require to ``rebox'' it by reintroducing it on the other side, while hard
reflecting walls require to invert the velocity component perpendicular to
the wall. After an event, be it a collision with another particle, a boundary
crossing or a reflection at a boundary, the event calendar must be re-evaluated
for pairs involving one of the event participants or a particle scheduled
to collide with one of the event participants. All other particles are not
influenced. Thus not every scheduled event actually takes place, because it can
be invalidated by another earlier event, in which case it is erased from the
priority queue. The latter is most commonly handled by means of a binary tree
\cite{Rapaport1980}, which we realized with a multimap of the C++ Standard
Template Library \cite{Josuttis1999}. The efficiency of this and alternative
data structures for event scheduling has been analyzed extensively
\cite{Marin1995,Paul2007}.

\begin{figure}[ht]
\includegraphics[width=\columnwidth]{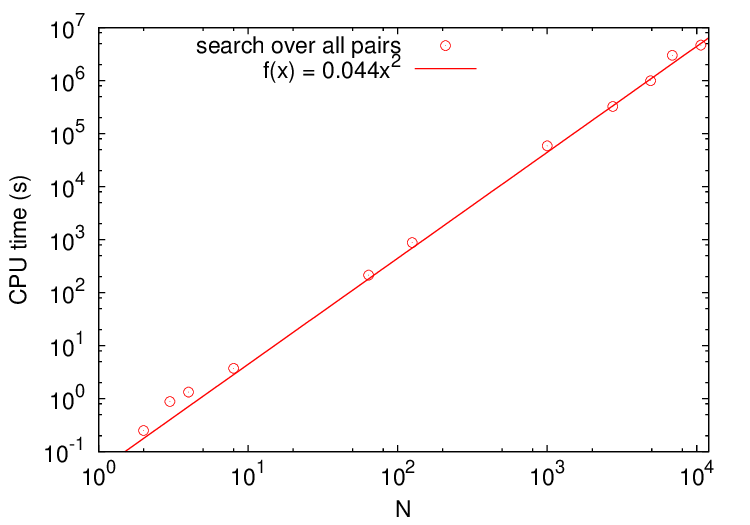}
\includegraphics[width=\columnwidth]{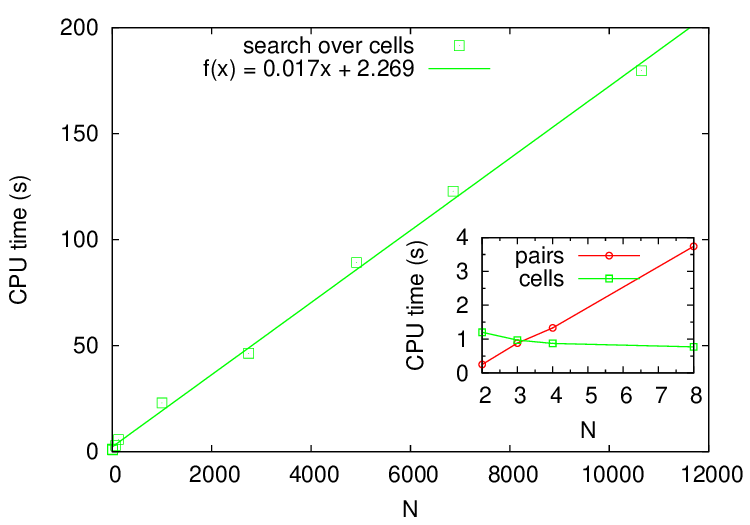}
\caption{\label{fig:benchmark_md}
(Color online) CPU time for $10^5$ collisions on a 2\,GHz Intel Core2 Duo as a
function of the number of hard spheres $N$ for our two event-based MD programs,
one with a simple search over all pairs (top), the other with an optimized
search over cells (bottom). The data are fitted respectively by a quadratic and
a linear function, which cross-over between $N = 3$ and $N = 4$ (inset). The
CPU times for the search over all pairs with $N \geq 1000$ were extrapolated
from the CPU times for 100 collisions.}
\end{figure}

The computational effort to search for $\min_{i,j} t_{ij}$ grows as the
square of the number of particles; see Fig.~\ref{fig:benchmark_md} (top).
For large systems it is advisable to divide the simulation box into cells
\cite{Erpenbeck1977}, which makes the dependence of the CPU time on the number
of particles linear; see Fig.~\ref{fig:benchmark_md} (bottom). Provided cell
boundary crossings are considered too in the event list, two particles can
collide only if they are located in the same cell or in adjacent cells.
We chose cells with a side larger than a particle diameter.
For more details on this and other algorithmic aspects in event-driven MD see
Refs.~\cite{Lubachevsky1991,Marin1993,Isobe1999}. For a parallel implementation
see Miller and Luding \cite{Miller2003}.

\clearpage

\bibliography{paper}

\end{document}